\documentclass[submission,copyright,creativecommons]{eptcs}
\providecommand{\event}{QPL 2022} 

\usepackage{iftex}

\ifpdf
 \usepackage{underscore}         
  \usepackage[T1]{fontenc}        
\else
\fi

\title{Logical Characterization of Contextual Hidden-Variable Theories 
Based on Quantum Set Theory
}
\author{Masanao Ozawa\thanks{Supported by JSPS KAKENHI Grant Numbers JP22K03424, JP21K11764, JP19H04066.}
\institute{Center for Mathematical Science and Artificial Intelligence, 
Academy of Emerging Sciences, Chubu University,\\ 
1200 Matsumoto-cho, Kasugai 487-8501, Japan}
\email{ozawa@isc.chubu.ac.jp}
\institute{Graduate School of Informatics, Nagoya University, Chikusa-ku, Nagoya 464-8601, Japan}
\email{ozawa@is.nagoya-u.ac.jp}
}
\date{}
\def\titlerunning{Hidden-Variable Theories and Quantum set theory}
\def\authorrunning{M. Ozawa}
\usepackage{graphicx}
\usepackage{amsthm}
\usepackage{mathptmx}
\usepackage{latexsym}
\usepackage{amsmath,amssymb,graphicx,color,bm}
\usepackage{enumerate}
\usepackage[nospace]{cite}
\usepackage{hyperref}
\usepackage{url}
\usepackage{doi}

\usepackage{bbm}
\usepackage[normalem]{ulem}
\renewcommand{\And}{\wedge}
\newcommand{\deq}[1]{ \begin{align}#1\end{align}}

\newcommand{\beq}{\begin{equation}}
\newcommand{\eeq}{\end{equation}}
\newcommand{\beqa}{\begin{eqnarray}}
\newcommand{\eeqa}{\end{eqnarray}}
\newtheorem{Theorem}{Theorem}[section]
\newcommand{\bTheorem}{\begin{Theorem}}
\newcommand{\eTheorem}{\end{Theorem}}
\newcommand{\C}{\mathbf{C}}
\newcommand{\Q}{\mathbbm{Q}}
\newcommand{\R}{\mathbbm{R}}
\newcommand{\al}{\alpha}
\newcommand{\be}{\beta}
\newcommand{\et}{\eta}
\newcommand{\la}{\lambda}
\newcommand{\mb}{\mbox}
\newcommand{\nn}{\nonumber}
\newcommand{\om}{\omega}
\newcommand{\ph}{\phi}
\newcommand{\ps}{\psi}
\newcommand{\rh}{\ps}
\newcommand{\si}{\sigma}
\newcommand{\vp}{\psi}
\newcommand{\vph}{\varphi}
\newcommand{\bx}{\mathbf{x}}
\newcommand{\De}{\Delta}
\newcommand{\DF}{{\cD}}  
\newcommand{\Ex}{{\rm Ex}}
\newcommand{\Iff}{\leftrightarrow}
\newcommand{\Inf}{\bigwedge}
\newcommand{\leo}{\le_{o}}
\newcommand{\Not}{\neg}
\newcommand{\Or}{\vee}

\newcommand{\Sup}{\bigvee}
\newcommand{\Then}{\rightarrow}

\newcommand{\dom}{{\rm dom}}
\newcommand{\bra}[1]{\langle#1|}
\newcommand{\bracket}[1]{\langle#1\rangle}
\newcommand{\ket}[1]{|#1\rangle}
\newcommand{\ketbra}[1]{\ket{#1}\bra{#1}}
\newcommand{\rank}{\mbox{\rm rank}}
\newcommand{\bL}{\mathbf{L}}
\newcommand{\bS}{\mathbf{S}}
\newcommand{\cA}{{\mathcal A}}
\newcommand{\cB}{{\mathcal B}}
\newcommand{\cC}{{\mathcal C}}
\newcommand{\cD}{{\mathcal D}}
\newcommand{\cH}{{\mathcal H}}
\newcommand{\cL}{{\mathcal L}}
\newcommand{\cM}{{\mathcal M}}
\newcommand{\cO}{{\mathcal O}}
\newcommand{\cP}{{\mathcal P}}
\newcommand{\cU}{{\mathcal U}}
\newcommand{\tA}{\tilde{A}}
\newcommand{\tX}{\tilde{X}}
\newcommand{\tY}{\tilde{Y}}
\newcommand{\forces}{\Vdash}
\newcommand{\val}[1]{[\![#1]\!]}
\newcommand{\vval}[1]{{[\![#1]\!]_{\cM}}}
\newcommand{\valo}[1]{[\![#1]\!]_{o}}
\newcommand{\cuniv}{\com}
\newcommand{\cm}{{\rm com}}
\newcommand{\com}{\mbox{{\rm com}}}
\newcommand{\p}{{}^{\perp}}
\newcommand{\hu}{\hat{u}}
\newcommand{\ck}[1]{\check{#1}}
\newcommand{\bitem}{\begin{itemize}\itemsep=0in}
\newcommand{\eenum}{\end{enumerate}}
\newcommand{\eitem}{\end{itemize}}
\newcommand{\benum}{\begin{enumerate}[{\rm (i)}]\itemsep=0in}
\newcommand{\BH}{\cL(\cH)}
\newcommand{\LL}{\bL_{0}}
\newcommand{\OM}{\cO(\cM)}
\newcommand{\PH}{\cP(\cH)}
\newcommand{\PM}{\cP(\cM)}
\newcommand{\PB}{\cP(\cB)}
\newcommand{\RM}{\R[\cM]}
\newcommand{\VB}{V[\cB]}
\newcommand{\VH}{V[\cH]}
\newcommand{\VM}{V[\cM]}

\begin{document}
\maketitle
\begin{abstract}
While non-contextual hidden-variable theories are proved to be impossible, contextual ones are possible.  In a contextual hidden-variable theory, an observable is called a beable if the hidden-variable assigns its value in a given measurement context specified by a state and a preferred observable.  Halvorson and Clifton characterized the algebraic structure of beables as a von Neumann subalgebra, called a beable subalgebra, of the full observable algebra such that the probability distribution of every observable affiliated therewith admits the ignorance interpretation.  On the other hand, we have shown that for every von Neumann algebra there is a unique set theoretical universe such that the internal ``real numbers'' bijectively correspond to the observables affiliated with the given von Neumann algebra. Here, we show that a set theoretical universe is associated with a beable subalgebra if and only if it is ZFC-satisfiable, namely, every theorem of ZFC set theory holds with probability equal to unity.  Moreover, we show that there is a unique maximal ZFC-satisfiable subuniverse ``implicitly definable'', in the sense of Malament and others,  by the given measurement context. The set theoretical language for the ZFC-satisfiable universe, characterized by the present work, rigorously reconstructs Bohr's notion of the ``classical language'' to describe the beables in a given measurement context.
\end{abstract}

\section{Introduction}
In 1935, Einstein, Podolsky, and Rosen \cite{EPR35} argued that the quantum mechanical description of physical reality is incomplete. Bohr \cite{Boh35} immediately responded to rebut their conclusion.  As to which claim is correct, the majority view has become in favor of Bohr, based on no-go theorems against non-contextual hidden-variable theories by von Neumann \cite{vN32E}, Gleason \cite{Gle57}, Kochen-Specker \cite{KS67}.  From the above debate it has been concluded that non-contextual hidden-variable theories for quantum mechanics are impossible. Nevertheless, contextual hidden-variable theories are possible as Bohr's complementarity interpretation \cite{Boh28} and the Bohmian mechanics \cite{Boh52}.  

In accordance with the above view, the modal interpretation of quantum mechanics has been studied extensively, used for no-collapse interpretation to solve the measurement problem \cite{Bub97}, and successfully articulated Bohr's otherwise obscure complementarity interpretation of quantum mechanics \cite{Bub97,HC02,12RBR}.   In modal interpretation, an observable possessing a well defined value is called a ``beable", and it is attempted to define the class of beables as broad as possible depending on the given measurement context.  The observable to be measured is naturally considered to possess its value to be revealed by the measurement even in a superposition of its eigenstates.  This avoids the measurement problem arising from state collapsing with the Dirac-von Neumann interpretation \cite{Dir58,vN32E} that an observable has a value only in its eigenstates or mixtures of them \cite{Bub97}.

Halvorson and Clifton \cite{HC99} algebraically characterized a contextual hidden-variable theory for a state $\ps$ in the Hilbert space $\cH$ of a quantum system $\bS$ as a von Neumann subalgebra $\cB$, called a beable subalgebra, of the full observable algebra $\BH$ such that the probability distribution of every observable therein admits the ignorance interpretation.  Their celebrated uniqueness theorem determines the unique maximal beable subalgebra $\cB(\ps,A)$ ``implicitly definable'', in the sense of Malament \cite{Mal77} and others, by the given measurement context $(\ps,A)$.  However, the algebraic structure of beables does not directly treat the logical structure of observational propositions nor the structure of the language speaking of beables, so that we do not have a formal framework to treat, for instance, Bohr's original notion of the ``classical language'' to describe beables in a given measurement context, or Hardy's logical formulation of non-locality.\footnote{We focus on the former topic in this paper and we will discuss the latter elsewhere.}

Here, we introduce a new approach based on quantum set theory to provide a logical framework for modal interpretations.  
Quantum set theory was introduced by Takeuti \cite{Ta81} for constructing mathematics based on quantum logic 
and developed by the present author \cite{07TPQ,14QST,16A2,17A1,17A2,21A6,21QTD}; the relationship with topos quantum theory was studied by Eva \cite{Eva15} and D\"{o}ring {\it et al.} \cite{21BBQ}.
In the preceding study, we have shown that for any von Neumann subalgebra $\cM$ of  the full observable algebra $\BH$ on a Hilbert space $\cH$, we can construct the unique mathematical, or more specifically, set theoretical, universe $\VM$ based on the logic represented by the projection lattice $\PM$ in $\cM$ such that the internal ``real numbers'' in the universe $\VM$  coincides with the self-adjoint operators (or observables) affiliated with the von Neumann algebra $\cM$.  

In this work, we shall logically characterize a contextual hidden-variable theory by showing that a von Neumann algebra $\cB$ is a beable subalgebra of $\BH$ for a state $\ps\in\cH$ if and only if the set theoretical universe $\VB$ based on the logic $\PB$ is ZFC-satisfiable in $\ps$, in the sense that every theorem of ZFC set theory in the language $\bL(\in,\VB)$ of set theory augmented by the names of elements of $\VB$ holds in $\VB$ with probability equal to unity in the state $\ps$.  In this case, the set of beables represented by self-adjoint operators affiliated with $\cB$ coincides with the set of the internal ``real numbers'' in the universe $\VB$.  Moreover, we uniquely determine the maximal ZFC-satisfiable subuniverse $V[\cB(\ps,A)]$ among those implicitly definable by the given measurement context $(\psi,A)$.  Thus, we can identify Bohr's notion of ``classical language'' describing the beables in the given measurement context $(\ps,A)$ with the language $\bL(\in,V[\cB(\ps,A)])$.\footnote{Bohr called the notion of beables in the given measurement context in several different ways, e.g., as the observables definable for the given measurement arrangement, or the elements of physical reality determined by the measurement of a preferred observable in the given state.}

\section{Algebraic approach to beables}
\label{se:UQ}
In this paper, we consider a quantum system $\bS$ described by a separable Hilbert space $\cH$, called the {\em state space} of $\bS$,
with inner product $(\xi,\et)$ for all $\xi,\et\in\cH$, linear in $\et$ and conjugate linear in $\xi$.
Observables of $\bS$ are bijectively represented by self-adjoint operators (densely
defined) on $\cH$ and every unit vector in $\cH$ represents a (pure) state of $\bS$.
If an observable (represented by a self-adjoint operator) $X$ is measured in a state (represented by a unit vector) $\ph\in\cH$,
the outcome $\bx$ of the measurement satisfies the Born statistical formula
\deq{
\Pr\{\bx\le x\|\ph\}=(\ph,E^{X}(x)\ph)
}  
for every real number $x\in \R$,
where $E^{X}(x)$ is the resolution of the identity belonging to 
the self-adjoint operator $X$ \cite[p.~119]{vN32E}.
We denote by $\mu^{X}_{\ph}$ the Borel probability measure on $\R$ uniquely determined by the relation
$\mu^{X}_{\ph}\left((-\infty,x]\right)=(\ph, E^{X}(x)\ph)$, and call it the {\em (Born) 
probability distribution} 
of the observable $X$ in the state $\ph$.
From the above,  for any (real-valued) Borel function $f$ the observable $f(X)$ defined by $f(X)=\int_{\R}f(x)dE^{X}(x)$ 
has the expectation value
\deq{
\Ex\{f(X)\|\ph\}=(\ph,f(X)\ph)
}
if $\ph\in\dom(f(X))$.

Now we consider a situation, called the {\em measurement context} $(\ps,A)$, in which an observable $A$ is to be measured in a state $\ps$, 
and we take the ignorance interpretation for the Born probability distribution $\mu^{A}_{\ps}$, as a typical reading of Bohr's complementarity interpretation \cite{Bub97,HC99,HC02,12RBR}, that just before the measurement of the observable $A$ in the state $\ps$, 
the observable $A$ possesses its value with the probability distribution 
$\mu^{A}_{\ps}$, and that the measurement faithfully reveals the value possessed by $A$.
We would call an observable that is considered to possess its value in the measurement context 
$(\ps,A)$ as a {\em beable} \cite[p.~41]{Bel73} in that measurement context.  
Obviously, the observable $A$ itself 
should be a beable together with its functions $f(A)$ for all Borel functions $f$.

The objective of modal interpretations is to determine the set of beables in the context $(\ps,A)$ as broad as possible.
We would call it the {\em maximal beable set}.
From the impossibility theorem of non-contextual hidden-variable theories by von Neumann \cite{vN32E} and others,\footnote{
The first general proof of the impossibility theorem for non-contextual hidden-variable theories was given by von Neumann \cite{vN32E}; see \cite{Bub10,Die17,MS18,Acu21} for the recent debate on the status of von Neumann's impossibility proof.
Later, Kochen and Specker \cite{KS67} proved the theorem for the Hilbert space  with the dimension greater
than 2 under the sole requirement that hidden-variables satisfy functional relations for observables;
a similar result can be derived as a corollary of Gleason's theorem \cite{Gle57}.
} 
the maximal beable set cannot be the whole set of observables.

It is natural to assume that the maximal beable set is closed under appropriate algebraic operations 
(i.e., addition, Jordan product, and Lie product) 
and closed under appropriate convergences, so that we assume that the maximal beable set is the set of observables (or self-adjoint operators) affiliated with a von Neumann algebra $\cB$ on 
$\cH$.\footnote{
An observable $X$ is affiliated with $\cB$ if and only if  
$E^{X}(x)\in\cB$ for all $x\in\R$. 
}

The first requirement for such a von Neumann algebra $\cB$ concerns the state $\ps$ requires that the Born probability distribution of every beable admit the ignorance interpretation.  
This requirement is mathematically formulated as follows. 
We call any normalized positive linear functional on $\cB$ a {\em state} on $\cB$.  A state $\om$ on $\cB$ is said to be 
{\em dispersion-free} iff $\om(X^*X)=|\om(X)|^2$ for any $X\in\cB$.
We say that a von Neumann algebra $\cB$ is {\em a beable subalgebra for a state vector $\psi\in\cH$} iff there is a probability measure $\mu$ on the space $\DF(\cB)$ of dispersion-free states of $\cB$ satisfying 
\deq{
(\ps,X\ps)=\int_{\DF(\cB)}\om(X)d\mu(\om)
} 
for every $X\in\cB$. 
The second requirement is, of course, that the observable $A$ be affiliated with $\cB$ as a ``privileged'' observable, 
from which it follows that $f(A)$ be affiliated with $\cB$ for all Borel functions $f$ by the Borel function calculus in von Neumann algebras.
Thus, we call a von Neumann algebra $\cB$ on $\cH$ a {\em beable subalgebra for the measurement context $(\vp,A)$} iff it satisfies following conditions:
\begin{enumerate}
\item[(i)]  (Beable) $\cB$ is a beable subalgebra for $\vp$.
\item[(ii)] ($A$-Priv) $A$ is affiliated with $\cB$.
\end{enumerate}

According to the above formulation, we are tempted to identity the maximal beable set with the set of
observables affiliated with a maximal von Neumann algebra among those satisfying the requirements ($A$-Priv)  
and (Beable) above.  However, such a choice is not unique.

To see this, consider the measurement context $(\psi,A)$ for the composite system $\bS$ of two spin 1/2 particles 
with the state space $\cH=\C^2\otimes\C^2$, consisting of the singlet state $\ps=2^{-1/2}(\ket{+_z}\ket{-_z}-\ket{-_z}\ket{+_z})$ 
and the $z$-component $A=\si_z\otimes I$ of Pauli spin operators of the first particle.
In this case, we have many beable subalgebras $\cB_{\theta}=W^*(\si_z)\otimes W^*(\si_\theta)$,
where $\si_\theta=\cos\theta \si_z+\sin \theta \si_x$, for $0\le \theta<\pi$.\footnote{For an observable $X$, 
we denote by $W^*(X)$ the von Neumann algebra generated by $X$ if $X$ is bounded, or the von Neumann
algebra generated by $E^{X}(x)$ for all $x\in\R$.}
Yet, there is no common maximal subalgebra $\cB_{\max}$, 
since if  $\cB_{\max}$ were to include $\cB_{\theta}$ and $\cB_{\theta'}$ with
$\theta\ne\theta'$, there would be no dispersion-free state on $\cB_{\max}$.

To consider which beable subalgebra we should choose, recall the debate between EPR \cite{EPR35} 
and Bohr \cite{Boh35} around the ``reality criterion'' proposed by EPR.
\begin{quote}
If, without in any way disturbing a system, we can predict with certainty (i.e., with probability equal to unity) the value of a physical quantity, 
then there exists an element of physical reality corresponding to this physical quantity.
\cite[p.777]{EPR35}
\end{quote}
EPR argued, in accordance with this criterion, that in the EPR state the position and momentum of the second particle have simultaneous reality, because the measurement on the first particle measures the second particle without disturbing it \cite[p.~334]{Sch35}.
Yet, the position and momentum cannot 
have simultaneous reality in any states by the uncertainty principle, so that EPR concluded that quantum-mechanical description of physical reality is incomplete. 
Bohr immediately responded to EPR.
 Bohr claimed that although the measurement on the first particle does not mechanically disturb the second particle,
the measurement on the first particle influences the condition that defines elements of reality for the second particle,  
and he rejected EPR's conclusion.
\begin{quote}
[T]here is in a case like that just considered no question of a mechanical disturbance of the system under investigation during the last critical stage of the measuring procedure. But even at this stage there is essentially the question of \textit{an influence on the very conditions which define the possible types of predictions regarding the future behavior of the system}. $\ldots$
[W]e see that the argumentation of the mentioned authors [EPR] does not justify their conclusion that quantum-mechanical description is essentially incomplete.''\cite[p.~700]{Boh35}
\end{quote}

Following the reality criterion posed by EPR \cite{EPR35}, but ``contextualized'' to the particular measurement context $(\psi,A)$
as suggested by Bohr \cite{Boh35} above, we should choose $B_0=I\otimes \si_z$ to be a beable in this measurement context 
but not $B_\theta=I\otimes \si_{\theta}$ with $\theta \ne 0$, because only the value of  $B_0$ can be inferred from 
the value of $A=\si_z\otimes I$ in the measurement context $(\psi,A)$ without disturbing the second particle.
If the observer were to measure the observable $A'=\si_x\otimes I$ instead of $A=\si_z\otimes I$, then from the value of $A'$ 
the observer could infer the value of $B_{\pi/2}=I\otimes \si_{x}$ without disturbing the second particle.  
EPR might have concluded that both $B_0$ and $B_{\pi/2}$ are beables, or elements of reality.
However, Bohr \cite{Boh35} pointed out that the status of being beable depends on the inference from the value of $A$ to the value of $B_0$ or the inference from the value of $A'$ to the value of $B_{\pi/2}$, but each inference is justified only in the
respective context, in which classical logic and classical mathematics can be used in the ordinary sense, 
and that there is no context-free classical language that supports the above two types of inferences simultaneously.
Thus, what are beables of the second particle depends on what is measured on the first particle as Bohr \cite{Boh35} suggested.

Halvorson and Clifton \cite[pp.14--15]{HC02} proposed a mathematical approach to single out appropriate beable
subalgebras consistent with the above ``contextualized'' reality criterion, as follows.
We say that a von Neumann algebra $\mathcal{B}$ is {\em implicitly
definable} by a measurement context $(\ps,A)$ 
iff $U^*\mathcal{B}U=\mathcal{B}$ for any unitary $U\in\cL(\cH)$
such that $U^*AU=A$ and $U\ps=\ps$.
In other words, $\cB$ is implicitly definable
by $(\vp,A)$ iff the membership relation $X\in\cB$ for any $X\in\cL(\cH)$ 
is not affected (i.e., $X\in\cB$ if and only if $\al(X)\in\cB$) 
by any automorphism $\al$ of $\cL(\cH)$ that leaves $\vp$ and $A$ invariant,\footnote{Any automorphism
$\al$ of $\cL(\cH)$ is of the form $\al(X)=U^{*}XU$ for some unitary $U\in\cL(\cH)$ \cite[p.~119]{Sak71}.}
which is widely used in foundational studies, for example, by Malament \cite[p.297]{Mal77}.
A beable subalgebra $\cB$ for $(\vp,A)$ is called {\em definable} iff it further satisfies 
\begin{enumerate}
\item[(iii)] (Def)  $\mathcal{B}$ is implicitly definable by $(\ps,A)$.
\end{enumerate}

We call a von Neumann algebra $\cB$ a {\em maximally definable beable subalgebra for the measurement
context $(\ps,A)$} iff $\cB$ satisfies ($A$-Priv), (Beable), and (Def) and  $\cB$ is maximal (in the set
inclusion) among all von Neumann subalgebras of $\BH$ satisfying those three requirements.
Then, the celebrated Halvorson-Clifton uniqueness theorem \cite{HC99} is stated as follows.

\bTheorem [Halvorson-Clifton \cite{HC99}]\label{th:HC02}
For any state $\psi$ and an observable $A$, there exists the unique
maximally definable beable subalgebra $\cB(\psi,A)$ for the measurement context 
$(\ps,A)$, and it is of the form
\deq{
\cB(\ps,A)=W^*(A)P\oplus \cL(P\p\cH),
}
where 
$P$ is the projection from $\cH$ onto the cyclic subspace $\cC(\ps,A)$ of $\cH$ generated by
$\ps$ and $A$, i.e, $\cC(\ps,A)=\{f(A)\psi\mid \mb{$f$ is a bounded Borel function}\}^{\perp\perp}$.
\eTheorem

\section{Quantum logic}

Bohr's view has, unfortunately, prevailed with several improper restatements, and there have been only a few serious attempts to reconstruct his reply in a rigorous analysis.
In his early contribution, Howard \cite{How79, How94} 
attempted to clarify what is the element of physical reality for Bohr.
He focused on Bohr's notion of ``classical description''.
In fact, Bohr emphasizes in several places that one should describe experimental evidence 
classically.\footnote{
``[I]t is decisive to recognize that, \textit{however far the phenomena transcend the scope of classical physical explanation, the account of all evidence must be expressed in classical terms}. The argument is simply that by the word ``experiment'' we refer to a situation where we can tell others what we have done and what we have learned and that, therefore, the account of the experimental arrangement and of the results of the  observations must be expressed in unambiguous language with suitable application of the terminology of classical physics.''\cite[p.209]{Boh49}
}

Howard's view was further sharpened by Halvorson and Clifton \cite{HC02} in the framework of modal interpretations as presented in the
preceding section; see also Ref.~\cite{12RBR} for mathematical refinements.
While Howard \cite{How79, How94} formulated Bohr's classicality requirement by the notion of ``appropriate'' mixture, Halvorson and Clifton reformulate it as the requirement that the Born probability distribution admit the ignorance interpretation. However, Bohr never explained his notion of ``classical description'' by the ignorance interpretation of the Born probability distribution.  In this paper, we attempt to go a step further.

In view of Bohr's writings, it seems that ``classical'' means classical physics, but he also stated:

\begin{quote}
[I]t would seem that the recourse to three-valued logic, sometimes proposed as means for dealing with the paradoxical features of quantum theory, is not suited to give a clearer account of the situation, since all well-defined experimental evidence, even if it cannot be analysed in terms of classical physics, must be expressed in ordinary language making use of common logic. \cite[p.~317]{Boh48}
\end{quote}

According to the above, we can see that what Bohr refers to in the word ``classical'' is a broader concept than classical physics and it is well understood as  the classical language that obeys classical logic rather than quantum logic.
Therefore, if the algebraic reconstruction should be consistent with Bohr's original view, the language that speaks of the beables should obey classical logic and theorems of classical mathematics.
The purpose of this paper is to rigorously realize this interpretation of Bohr's view 
in the framework of quantum set theory.

Recall that the logic of observational propositions on the system $\bS$ is represented by the projection lattice $\PH$ of all projections on $\cH$.
In order to treat a context-dependent part of the whole observational propositions, we consider a sublogics of $\PH$ represented by the projection lattice $\PM$ of  a von Neumann subalgebra $\cM$ of $\BH$.
Let $\cM$ be a von Neumann algebra on $\cH$ and denote by $\OM$ the set of observables affiliated with $\cM$.
The observational propositions on the observables affiliated with $\cM$ are constructed from {\em atomic propositions} ``$X\leo \la$'', where $X\in\OM$ and $\la\in\R$, by connecting them with $\And$ (conjunction),
$\Or$ (disjunction), $\Then$ (conditional), $\Iff$ (equivalence), and $\Not$ (negation). 
In what follows, we consider only the conjunction and negation as primitive symbols and the other connectives as derived
symbols by the following  definitions: $\vph_1\Or\vph_2=\Not(\Not\vph_1\And\Not\vph_2)$, 
$\vph_1\Then\vph_2=\Not\vph_1\Or(\vph_1\And\vph_2)$, 
$\vph_1\Iff\vph_2=(\vph_1\Then\vph_2)\And(\vph_2\Then\vph_1)$.
Here, we note that the conditional $\Then$ is defined as the Sasaki conditional \cite{Sas54}.
We denote by $\bL(\cM)$ the set of observational propositions on the observables in $\OM$.
We also write $\bL(\cH)=\bL(\BH)$.
We have $\bL(\cM_2)\subseteq\bL(\cM_2)$ if $\cM_1\subseteq\cM_2$.

Following Birkhoff and von Neumann \cite{BvN36},
every observational proposition $\vph$ has the projection-valued truth value 
$\valo{\vph}$ determined by the following rules.

(i) $\valo{X\leo \la}=E^{X}(\la)$.

 (ii) $\valo{\vph_1\And\vph_2}=\valo{\vph_1}\And \valo{\vph_2}$.
 
 (iii) $\valo{\Not\vph}=\valo{\vph}\p$.
 
 Then for every observational proposition $\vph$ we define the probability that the observational proposition 
 $\vph$ holds in the state $\psi$ by $\Pr\{\vph\|\psi\}=\|\valo{\vph}\psi\|^2$.
 It is well-know that the logic of observational propositions in $\bL(\cH)$ is non-distributive,  
 so that it does not necessarily follow the laws of classical logic.
\sloppy
According to Bohr's view on ``classical description'' it is natural to expect that in the state $\ps$ the language $\bL(\cB)$ satisfies classical logic and all the mathematical theorems, if $\cB$ is beable for the state $\psi$.
However, the language $\bL(\cM)$ has only a limited power in expressing observational relations between observables,
just as the propositional logic has only a limited power in expressing relations between mathematical objects.  
In fact, the language $\bL(\cM)$ cannot generally express relations between observables in $\OM$ 
such as equality and order relation,  so that it cannot assign the projection valued truth-value, or the probability
in the state $\vp$, for the proposition ``A and B have the same value''.
In what follows we strengthen the language of observational propositions to have full power of expressing all the
mathematically definable relations between observables with projection-valued truth values. 

\section{Quantum set theory}

In this and the next sections, we shall introduce basic facts about quantum set theory.
We refer the reader to Ref.~\cite{21QTD} for detailed formulations.
Let $\cM$ be a von Neumann algebra.
The purpose of quantum set theory is to extend the universe, or a ground model, $V$ of ZFC set theory to 
a sort of ``generic extension'' $V[\cM]$ adding self-adjoint operators affiliated with $\cM$ as ``generic
reals'' to $V$.
 
We denote by $V$ the universe of the Zermelo-Fraenkel set theory
with the axiom of choice (ZFC).
Let $\bL(\in)$ be the language for first-order theory with equality 
augmented by a binary relation symbol
$\in$, bounded quantifier symbols $\forall x\in y$, $\exists x \in y$, and no constant symbols.
For any class $U$, the language $\bL(\in,U)$ is the one obtained by adding a name 
for each element of $U$.
We take the symbols $\Not$, $\And$, $\forall x\in y$, and $\forall x$
as primitive, and the symbols $\Or$, $\Then$, $\Iff$, 
$\exists x\in y$, and $\exists x$ as derived
symbols by defining:
\benum
\item $\vph_1\Or\vph_2=\Not(\Not\vph_1\And\Not\vph_2)$, 
\item $\vph_1\Then\vph_2=\Not\vph_1\Or(\vph_1\And\vph_2)$, 
\item $\vph_1\Iff\vph_2=(\vph_1\Then\vph_2)\And(\vph_2\Then\vph_1)$,
\item $\exists x\in y\,\vph(x)=\Not(\forall x\in y\,\Not\vph(x)),$
\item $\exists x\vph(x)=\Not(\forall x\,\Not\vph(x)).$
\eenum

Let $\cM$ be a von Neumann algebra on a Hilbert space $\cH$.
For each ordinal $ {\al}$, let
\beq
V_{\al}[\cM] = \left\{u\left|\ u:\dom(u)\to \PM \mbox{ and }
(\exists \be<\al)\ 
\dom(u) \subseteq V_{\be}[\cM]\right.\right\}.\footnote{We denote by $\dom(f)$ the domain of a function $f$.
By $f:D\to R$ we mean that $f$ is a function defined on a set $D$  with values in a set $R$.} 
\eeq
The {\em $\PM$-valued universe} $\VM
 $ is defined
by 
\beq
  \VM
 = \bigcup _{{\al}{\in}\mbox{On}} V_{{\al}}[\cM],
\eeq
where $\mbox{On}$ is the class of all ordinals. 
We shall write $V_\al[\cH]=V_\al[\cL(\cH)]$ and $\VH=V[\cL(\cH)]$.
For every $u\in\VM$, the {\em rank} of $u$, denoted by
$\rank(u)$,  is defined as the least $\al$ such that
$u\in V_{\al+1}[\cM]$.
It is easy to see that if $u\in\dom(v)$ then 
$\rank(u)<\rank(v)$.

We introduce the implication operation $\Then$ and its dual conjunction operation $*$ on the lattice
$\PM$ by $P\Then Q=P\p\Or(P\And Q)$ and $P*Q=P\And(P\p\Or Q)$, or equivalently $P*Q=(P\Then Q\p)\p$, for any $P,Q\in\PM$.
The operation $\Then$ on $\PM$ is called the Sasaki arrow and the operation $Q\mapsto P*Q$ is called
the Sasaki projection \cite{Kal83}.

For any $u,v\in\VM
 $, the $\PM$-valued truth values of
atomic formulas $u=v$ and $u\in v$ are assigned
by the following rules recursive in rank.
\begin{enumerate}[(i)]\itemsep=0in
\setcounter{enumi}{5}
\item $\vval{u = v}
= \displaystyle{\inf_{u' \in  \dom(u)}}(u(u') \Then
\vval{u'  \in v})
\And \displaystyle{\inf_{v' \in   \dom(v)}}(v(v') 
\Then  \vval{v'  \in u})$.
\item $\displaystyle{ \vval{u \in v} 
= \sup_{v' \in \dom(v)} (v(v')*\vval{u =v'})}$.
\end{enumerate}

To each statement $\vph$ of $\bL(\in,\VM
 )$ 
we assign the $\PM$-valued truth value $ {\vval{\vph}}$ by the following
rules.
\begin{enumerate}[(i)]\itemsep=0in
\setcounter{enumi}{7}
\item $ \vval{\Not\vph} = \vval{\vph}^{\perp}$.
\item $ \vval{\vph_1\And\vph_2} 
= \vval{\vph_{1}} \And \vval{\vph_{2}}$.
\item $ \vval{(\forall x\in u)\, {\vph}(x)} 
= \displaystyle{\Inf_{u'\in \dom(u)}
(u(u') \Then
  \vval{\vph(u')})}$.
\item $\displaystyle{ \vval{(\forall x)\, {\vph}(x)} 
= \Inf_{u\in \VM
 }\vval{\vph(u)}}$.
\end{enumerate}

We say that a statement ${\vph}$ of $ \bL(\in,\VM
 ) $
{\em holds} in $\VM
 $ if $ \vval{{\vph}} = I$.
A formula in $\bL(\in,\VM)$ is called a {\em
$\De_{0}$-formula}  iff it has no unbounded quantifiers
$\forall x$ or $\exists x$.
The following theorem holds \cite[Theorem 4.3]{21QTD}.

\sloppy
\begin{Theorem}[$\De_{0}$-Absoluteness Principle]
\label{th:Absoluteness}
\sloppy  
For any $\De_{0}$-formula 
${\vph} (x_{1},{\ldots}, x_{n}) $ 
of $\bL(\in)$ and $u_{1},{\ldots}, u_{n}\in \VM
 $, 
we have
\deq{
\vval{\vph(u_{1},\ldots,u_{n})}=
{\val{\vph(u_{1},\ldots,u_{n})}}_{\BH}.
}
\end{Theorem}

Henceforth, 
for any $\De_{0}$-formula 
${\vph} (x_{1},{\ldots}, x_{n}) $
and $u_1,\ldots,u_n\in\VM
 $,
we abbreviate ${\val{\vph(u_{1},\ldots,u_{n})}}=
\vval{\vph(u_{1},\ldots,u_{n})}$,
which is the common $\PH$-valued truth value for 
$u_{1},\ldots,u_{n}\in\VH$.

The universe $V$  can be embedded in
$\VM
 $ by the following operation 
$\vee:v\mapsto\check{v}$ 
defined by the $\in$-recursion: 
for each $v\in V$, $\check{v} = \{\check{u}|\ u\in v\} 
\times \{I\}$. 
Note that $\check{v}\in\VM$ for any $v\in V$ and any von Neumann subalgebra $\cM\subseteq\BH$.
Then we have the following \cite[Theorem 4.8]{21QTD}.

\begin{Theorem}[$\De_0$-Elementary Equivalence Principle]
\label{th:2.3.2}
\sloppy  
For any $\De_{0}$-formula 
${\vph} (x_{1},{\ldots}, x_{n}) $ 
of $\bL(\in)$ and $u_{1},{\ldots}, u_{n}\in V$,
we have
$
\bracket{V,\in}\models {\vph}(u_{1},{\ldots},u_{n})
\mbox{ if and only if }
\val{\vph(\check{u}_{1},\ldots,\check{u}_{n})}=I.
$
\end{Theorem}

Thus, $\VM$ includes (a copy of) the standard universe $V$ as a $\De_0$-elementary equivalent
submodel.
For further detail about the universe $\VM$ we refer the reader to \cite{21QTD}.

\section{Transfer principle}
\label{se:CIQL}

For any $u\in\VM
 $, we define the {\em support} of $u$, denoted by $L(u)$, by transfinite recursion on the 
rank of $u$ by the relation
\beq
L(u)=\bigcup_{x\in\dom(u)}L(x)\cup\{u(x)\mid x\in\dom(u)\}\cup\{0\}.
\eeq
For $\cA\subseteq\VM
 $ we write 
$L(\cA)=\bigcup_{u\in\cA}L(u)$ and
for $u_1,\ldots,u_n\in\VM
 $ we write 
$L(u_1,\ldots,u_n)=L(\{u_1,\ldots,u_n\})$.
Let $\cA\subseteq\BH$. 
The {\em commutant of $\cA$ in $\BH$}, denoted by $\cA'$, is defined by
\deq{
\cA'&=\{A\in\BH\mid \mb{$[A,B]=0$  for all $B\in\cA$}\},
}
and the {\em commutant of $\cA$ in $\PH$}, denoted by $\cA^{!}$, is defined by
$\cA^{!}=\cA'\cap\PH$.

Let $\cA\subseteq\PH$.  
Takeuti \cite{Ta81} introduced the {\em commutator} of $\cA$, denoted by  $\com(\cA)$, given by
\deq{
\com(\cA)=\Sup\{E\in\cA^{!}\cap\cA^{!!}\mid \mb{$[P,Q]E=0$ for all $P,Q\in\cA$}\}.
}  
For any $P_1,\ldots,P_n\in\PH$, we write $\com(\{P_1,\ldots,P_n\})=\com(P_1,\ldots,P_n)$.
We refer the reader to \cite{16A2} for further properties of commutators.

Let $\cA\subseteq\VM
 $.  The {\em commutator
of $\cA$}, denoted by $\cm(\cA)$, is defined by 
\beq
\cuniv(\cA)=\com (L(\cA)).
\eeq
For any $u_1,\ldots,u_n\in\VM
 $, we write
$\cuniv(u_1,\ldots,u_n)=\cuniv(\{u_1,\ldots,u_n\})$.

We have the following transfer principle for bounded theorems of ZFC
\cite[Theorem 4.15]{21QTD}.

\begin{Theorem}[$\De_{0}$-ZFC  Transfer Principle]
For any $\De_{0}$-formula ${\vph} (x_{1},{\ldots}, x_{n})$ 
of $\bL(\in)$ and $u_{1},{\ldots}, u_{n}\in\VM$, if 
${\vph} (x_{1},{\ldots}, x_{n})$ is provable in ZFC, then
\beq
\val{\vph({u}_{1},\ldots,{u}_{n})}
\ge
\cuniv(u_{1},\ldots,u_{n}).
\eeq
\end{Theorem}

\section{Internal real numbers in quantum set theory}
\label{se:RN}

Let $\Q$ be the set of rational numbers in $V$.
We define the set of rational numbers in the model $\VM
 $
to be $\check{\Q}$.
We define a real number in the model by a Dedekind cut
of the rational numbers. More precisely, we identify
a real number with the upper segment of a Dedekind cut whose
lower segment has no end point.
Therefore, the formal definition of  the predicate $\R(x)$, 
``$x$ is a real number,'' is expressed by
\beqa
\R(x)&:=&
\forall y\in x(y\in\check{\Q})
\And \exists y\in\check{\Q}(y\in x)
\And \exists y\in\check{\Q}(y\not\in x)\nn\\
& &  \And
\forall y\in\check{\Q}(y\in x\Iff\forall z\in\check{\Q}
(y<z \Then z\in x)).
\eeqa
We define $\R[\cM]$ 
to be the interpretation of 
the set $\R$ of real numbers in $\VM$ 
as follows.
\deq{
\R[\cM] = \{u\in\VM
 \mid \dom(u)=\dom(\check{\Q})
\ \mb{and }\val{\R(u)}=I\}.
}

For any $u\in\R[\cM]$ and $\la\in\R$, we define $E^{u}(\la)$ by 
\beq
E^{u}(\la)=\Inf_{\la<r\in\Q}u(\check{r}).
\eeq
Then it can be shown that
$\{E^u(\la)\}_{\la\in\R}$ is a resolution of
identity in $\PM$ and hence by the spectral theorem there
is an observable $\hat{u}\in\OM$ uniquely
satisfying $\hat{u}=\int_{\R}\la\, dE^u(\la)$.  On the other
hand, let $A\in\OM$. We define $\tilde{A}\in\VM
 $ by
\beq
\tilde{A}=\{(\check{r},E^{A}(r))\mid r\in\Q\}.
\eeq
Then $\dom(\tA)=\dom(\check{\Q})$ and $\tA(\check{r})=E^{A}(r)$ for all $r\in\Q$.
It is easy to see that $\tilde{A}\in\RM$ and we have
$(\hat{u})\tilde{\ }=u$ for all $u\in\RM$ and $(\tilde{A})\hat{\ }=A$
for all $A\in\OM$.
Therefore, the correspondence
between $\RM$ and $\OM$ is bijective.
We call the above correspondence the {\em Takeuti correspondence}.
Now, we have the following.

\bTheorem
The relations 
\bitem
\item[\rm (i)] ${\displaystyle E^{A}(\la)=\Inf_{\la<r\in\Q}u(\check{r})}$\quad  for all $\la\in\Q
$,
\item[\rm (ii)] $u(\check{r})=E^{A}(r)$\quad  for all $r\in\Q$,
\eitem
where $u=\tA\in\RM$ and $A=\hu\in \OM$, 
sets up a bijective correspondence between $u\in\RM$ and $A\in\OM$.
\eTheorem

For any $r\in\R$, we shall write $\tilde{r}=(r1)\,\tilde{}$, where $r1$ is the scalar 
operator on $\cH$.
Then we have $\dom(\tilde{r})=\dom(\check{\Q})$ and $\tilde{r}(\check{t})=
\val{\check{r}\le \check{t}}$, so that we have $L(\tilde{r})=\{0,1\}$.
The order relation for $u,v\in\RM$ is naturally defined by
\deq{
u\le v := (\forall x\in\ck{\Q})\ [x\in v \Then x\in u].
}

Recall that a formula $\vph(x_1,\ldots,x_n)\in\bL(\in,\VM)$ is called a $\De_0$-formula iff it contains
no unbounded quantifiers $\forall x$ nor $\exists x$.
In this paper, we focus on the sublanguage $\LL(\in,\VM)$ consisting of $\De_0$-formulas
in $\bL(\in,\VM)$.
Then, for every statement $\vph\in\LL(\in,\VM)$ we have the $\PM$-valued truth value 
$\val{\vph}=\val{\vph}_{\BH}$,
and for every observable $X\in\OM$ we have a real number $\tX\in\RM$ in $\VM$.
Thus, there is an embedding of every observational proposition $\vph$ 
in the language $\bL(\cM)$ into a statement $\tilde{\vph}$ in
the language $\LL(\in,\VM)$ defined by the following rules for any 
$X\in\OM$ and $x\in\R$, 
and observational propositions $\vph,\vph_1,\vph_2$:
\bitem
\item[(Q1)] $\displaystyle\widetilde{X\leo x}:=\tilde{X}\le\tilde{x}$.
\item[(Q2)] $\widetilde{\Not \vph}:=\Not\tilde{\vph}.$
\item[(Q3)] $\widetilde{\vph_1\And \vph_2}:={\tilde\vph_1}\And\tilde{\vph_2}.$
\eitem
Then, it is easy to see that the relation
\beq
\val{\tilde{\vph}}=\valo{\vph}
\eeq
holds for any observational proposition $\vph$.  

Thus, all the observational propositions are embedded in a set of 
statements in $\LL(\in,\VM)$ preserving their projection-valued truth values.
Quantum set theory provides a language strong enough to express 
all the possible mathematical relations among observables.  
For every mathematical relation $R(x_1,\ldots,x_n)$ definable in ZFC set theory,
we can assign the quantum logical truth value $\val{R(\tX_1,\ldots,\tX_n)}\in\PM$ 
of the relation $R(\tX_1,\ldots,\tX_n)$ 
for any observables $X_1,\ldots,X_n\in\cO(\cH)$.
For example, we can generally define the projection valued truth values for the  
order relation and the equality relation for observables so that the following relation hold:
\deq{
\val{\tX\le\tY}&=\val{(\forall x\in\ck{\Q})\ [\tY\le x\Then \tX\le x]},\\
\val{\tX=\tY}&=\val{\tX\le \tY \And \tY\le\tX}.
}

\section{Beable subuniverses} 
Let $\ps\in\cH$ be a state. We say that 
a statement $\vph(u_1,\ldots,u_n)\in\LL(\in,\VM)$ {\em holds 
in $\ps$}, and write $\ps\forces\vph(u_1,\ldots,u_n)$, iff $\Pr\{\vph(u_1,\ldots,u_n)\| \ps\}=I$. 
For any von Neumann algebra
 $\cM\subseteq\BH$, 
the subuniverse $\VM
 \subseteq\VH$ is said to be {\em ZFC-satisfiable in $\ps$} iff
\deq{
\rh\forces\ph(u_1,\ldots,u_n)
}
for any $u_1,\ldots,u_n\in\VM$ and any formula $\vph(x_1,\ldots,x_n)\in\LL(\in)$
provable in ZFC.

The following theorem holds.

\begin{Theorem}\label{th:201225}
Let $\cM$ be a von Neumann algebra on $\cH$.
Then the following conditions are all equivalent.
\benum
\item  $\VM
 $ is ZFC-satisfiable in $\ps$.
\item $\cM$ is beable for $\rh$.
\item $[X,Y]\ps=0$ for any $X,Y\in\cM$.
\item $\ketbra{\ps}\le \cm(u_1,\ldots,u_n)$ for any $u_1,\ldots,u_n\in\VM$.
\eenum
\end{Theorem}

\section{Context-definable beable subuniverses}
Let $(\ps,A)$ be a measurement context.
For any unitary operator $U$ on $\cH$, we define $\al_{U}:\VH\to\VH$
by transfinite recursion on the rank of $u\in\VH$ 
as
\deq{
\al_{U}(u)=\{\bracket{\al_{U}(x),U^{*} u(x) U} \mid x\in\dom(u)\}.
}

A subclass $\cU\subseteq\VM
 $ is called {\em definable} by $(\ps,A)$ iff

(i) $\tilde{A}\in\cU$,

(ii)  $\al_{U}(\cU) \subseteq \cU$ for any unitary operator $U$ on $\cH$ satisfying
$[A,U]=0$ and $U\ps=\ps$.

A subuniverse $\VM\subseteq\VH$ is called a {\em maximally definable ZFC-satisfiable subuniverse} for
the measurement context $(\ps,A)$ iff $\VM$ is a ZFC-satisfiable subuniverse in $\ps$ definable by $(\ps,A)$
and there is no ZFC-satisfiable subuniverse $V[\cM_0]$ definable by $(\ps,A)$ that properly includes $\VM$.
Then, we have the following.

\begin{Theorem}
(i) A subuniverse $\VM$ is a ZFC-satisfiable subuniverse of $\VH$ definable by $(\ps,A)$ if and only
if $\cM$ is a beable subalgebra of $\BH$ definable by $(\ps,A)$.

(ii) There uniquely exists a maximally definable ZFC-satisfiable subuniverse $\VM\subseteq\VH$ for 
any measurement context $(\ps,A)$. In this case, $\cM$ is of the form $\cM=\cB(\ps,A)$, i.e.,
\deq{
\cM=W^*(A)P\oplus \cL(P\p\cH),
}
where 
$P$ is the projection from $\cH$ onto the cyclic subspace $\cC(\ps,A)$ of $\cH$ generated by
$\ps$ and $A$.
\end{Theorem}


\begin{thebibliography}{10}
\providecommand{\bibitemdeclare}[2]{}
\providecommand{\surnamestart}{}
\providecommand{\surnameend}{}
\providecommand{\urlprefix}{Available at }
\providecommand{\url}[1]{\texttt{#1}}
\providecommand{\href}[2]{\texttt{#2}}
\providecommand{\urlalt}[2]{\href{#1}{#2}}
\providecommand{\doi}[1]{doi:\urlalt{https://doi.org/#1}{#1}}
\providecommand{\eprint}[1]{arXiv:\urlalt{https://arxiv.org/abs/#1}{#1}}
\providecommand{\bibinfo}[2]{#2}

\bibitemdeclare{article}{Acu21}
\bibitem{Acu21}
\bibinfo{author}{P.~\surnamestart Acu{\~{n}}a\surnameend}
  (\bibinfo{year}{2021}): \emph{\bibinfo{title}{Von {Neumann's} Theorem
  Revisited}}.
\newblock {\slshape \bibinfo{journal}{Found.\ Phys.}} \bibinfo{volume}{51}, pp.
  \bibinfo{pages}{73/1--73/29}, \doi{10.1007/s10701-021-00474-5}.

\bibitemdeclare{incollection}{Bel73}
\bibitem{Bel73}
\bibinfo{author}{J.~S. \surnamestart Bell\surnameend} (\bibinfo{year}{2004}):
  \emph{\bibinfo{title}{Subject and object}}.
\newblock In: {\slshape \bibinfo{booktitle}{Speakable and Unspeakable in
  Quantum Mechanics: Collected Papers on Quantum Philosophy}},
  \bibinfo{edition}{2nd} edition, \bibinfo{publisher}{Cambridge UP}, pp.
  \bibinfo{pages}{40--44}, \doi{10.1017/CBO9780511815676.007}.

\bibitemdeclare{article}{BvN36}
\bibitem{BvN36}
\bibinfo{author}{G.~\surnamestart Birkhoff\surnameend} \&
  \bibinfo{author}{J.~\surnamestart von Neumann\surnameend}
  (\bibinfo{year}{1936}): \emph{\bibinfo{title}{The Logic of Quantum
  Mechanics}}.
\newblock {\slshape \bibinfo{journal}{Ann.\ Math.}} \bibinfo{volume}{{37}}, pp.
  \bibinfo{pages}{823--843}, \doi{10.2307/1968621}.

\bibitemdeclare{article}{Boh52}
\bibitem{Boh52}
\bibinfo{author}{D.~\surnamestart Bohm\surnameend} (\bibinfo{year}{1952}):
  \emph{\bibinfo{title}{A Suggested Interpretation of the Quantum Theory in
  Terms of ``{Hidden Variables},'' {I}, {II}}}.
\newblock {\slshape \bibinfo{journal}{Phys,\ Rev.}} \bibinfo{volume}{85}, pp.
  \bibinfo{pages}{166--179, 180--193}, \doi{10.1103/PhysRev.85.166,
  10.1103/PhysRev.85.180}.

\bibitemdeclare{article}{Boh28}
\bibitem{Boh28}
\bibinfo{author}{N.~\surnamestart Bohr\surnameend} (\bibinfo{year}{1928}):
  \emph{\bibinfo{title}{The Quantum Postulate and the Recent Development of
  Atomic Theory}}.
\newblock {\slshape \bibinfo{journal}{Nature}} \bibinfo{volume}{{121}}, pp.
  \bibinfo{pages}{580--590}, \doi{10.1038/121580a0}.

\bibitemdeclare{article}{Boh35}
\bibitem{Boh35}
\bibinfo{author}{N.~\surnamestart Bohr\surnameend} (\bibinfo{year}{1935}):
  \emph{\bibinfo{title}{Can Quantum-Mechanical Description of Physical Reality
  be Considered Complete?}}
\newblock {\slshape \bibinfo{journal}{Phys.\ Rev.}} \bibinfo{volume}{48}, pp.
  \bibinfo{pages}{696--702}, \doi{10.1103/PhysRev.48.696}.

\bibitemdeclare{article}{Boh48}
\bibitem{Boh48}
\bibinfo{author}{N.~\surnamestart Bohr\surnameend} (\bibinfo{year}{1948}):
  \emph{\bibinfo{title}{On the Notions of Causality and Complementarity}}.
\newblock {\slshape \bibinfo{journal}{Dialectica}} \bibinfo{volume}{2}, pp.
  \bibinfo{pages}{312--319}, \doi{10.1111/j.1746-8361.1948.tb00703.x}.

\bibitemdeclare{incollection}{Boh49}
\bibitem{Boh49}
\bibinfo{author}{N.~\surnamestart Bohr\surnameend} (\bibinfo{year}{1949}):
  \emph{\bibinfo{title}{Discussion with {Einstein} on epistemological problems
  in atomic physics}}.
\newblock In \bibinfo{editor}{P.~A. \surnamestart Shilpp\surnameend}, editor:
  {\slshape \bibinfo{booktitle}{Albert Einstein: Philosopher-Scientist}},
  {\slshape \bibinfo{series}{The Library of Living Philosophers}}
  \bibinfo{volume}{VII}, \bibinfo{publisher}{Northwestern University},
  \bibinfo{address}{Evanston}, pp. \bibinfo{pages}{200--241},
  \doi{10.1016/S1876-0503(08)70379-7}.

\bibitemdeclare{book}{Bub97}
\bibitem{Bub97}
\bibinfo{author}{J.~\surnamestart Bub\surnameend} (\bibinfo{year}{1997}):
  \emph{\bibinfo{title}{Interpreting the Quantum World}}.
\newblock \bibinfo{publisher}{Cambridge UP}, \bibinfo{address}{Cambridge}.

\bibitemdeclare{article}{Bub10}
\bibitem{Bub10}
\bibinfo{author}{J.~\surnamestart Bub\surnameend} (\bibinfo{year}{2010}):
  \emph{\bibinfo{title}{Von {Neumann's} `No Hidden Variables' Proof: {A}
  Re-Appraisal}}.
\newblock {\slshape \bibinfo{journal}{Found.\ Phys.}} \bibinfo{volume}{40}, pp.
  \bibinfo{pages}{1333--1340}, \doi{10.1007/s10701-010-9480-9}.

\bibitemdeclare{article}{Die17}
\bibitem{Die17}
\bibinfo{author}{D.~\surnamestart Dieks\surnameend} (\bibinfo{year}{2017}):
  \emph{\bibinfo{title}{Von {Neumann's} impossibility proof: {Mathematics} in
  the service of rhetorics}}.
\newblock {\slshape \bibinfo{journal}{Stud.\ Hist.\ Philos.\ Sci.\ B}}
  \bibinfo{volume}{60}, pp. \bibinfo{pages}{136--148},
  \doi{10.1016/j.shpsb.2017.01.008}.

\bibitemdeclare{book}{Dir58}
\bibitem{Dir58}
\bibinfo{author}{P.~A.~M. \surnamestart Dirac\surnameend}
  (\bibinfo{year}{1958}): \emph{\bibinfo{title}{The Principles of Quantum
  Mechanics}}, \bibinfo{edition}{4th} edition.
\newblock \bibinfo{publisher}{Oxford UP}, \bibinfo{address}{Oxford},
\doi{10.1063/1.3062610}.

\bibitemdeclare{article}{21BBQ}
\bibitem{21BBQ}
\bibinfo{author}{A.~\surnamestart D\"{o}ring\surnameend},
  \bibinfo{author}{B.~\surnamestart Eva\surnameend} \&
  \bibinfo{author}{M.~\surnamestart Ozawa\surnameend} (\bibinfo{year}{2021}):
  \emph{\bibinfo{title}{A Bridge Between Q-Worlds}}.
\newblock {\slshape \bibinfo{journal}{Rev.~Symb.~Log.}}
  \bibinfo{volume}{14}(\bibinfo{number}{2}), pp. \bibinfo{pages}{447--486},
  \doi{10.1017/S1755020319000492}.

\bibitemdeclare{article}{EPR35}
\bibitem{EPR35}
\bibinfo{author}{A.~\surnamestart Einstein\surnameend},
  \bibinfo{author}{B.~\surnamestart Podolsky\surnameend} \&
  \bibinfo{author}{N.~\surnamestart Rosen\surnameend} (\bibinfo{year}{1935}):
  \emph{\bibinfo{title}{Can Quantum-Mechanical Description of Physical Reality
  be Considered Complete?}}
\newblock {\slshape \bibinfo{journal}{Phys.\ Rev.}} \bibinfo{volume}{{47}}, pp.
  \bibinfo{pages}{777--780}, \doi{10.1103/PhysRev.47.777}.

\bibitemdeclare{article}{Eva15}
\bibitem{Eva15}
\bibinfo{author}{B.~\surnamestart Eva\surnameend} (\bibinfo{year}{2015}):
  \emph{\bibinfo{title}{Towards a Paraconsistent Quantum Set Theory}}.
\newblock {\slshape \bibinfo{journal}{Electronic Proceedings in Theoretical
  Computer Science}} \bibinfo{volume}{195}, pp. \bibinfo{pages}{158--169},
  \doi{10.4204/EPTCS.195.12}.

\bibitemdeclare{article}{Gle57}
\bibitem{Gle57}
\bibinfo{author}{A.~M. \surnamestart Gleason\surnameend}
  (\bibinfo{year}{1957}): \emph{\bibinfo{title}{Measures on the Closed
  Subspaces of a {Hilbert} Space}}.
\newblock {\slshape \bibinfo{journal}{J. Math.\ Mech.}} \bibinfo{volume}{{6}},
  pp. \bibinfo{pages}{885--893}, \doi{10.1512/iumj.1957.6.56050}.

\bibitemdeclare{article}{HC99}
\bibitem{HC99}
\bibinfo{author}{H.~\surnamestart Halvorson\surnameend} \&
  \bibinfo{author}{R.~\surnamestart Clifton\surnameend} (\bibinfo{year}{1999}):
  \emph{\bibinfo{title}{Maximal Beable Subalgebras of Quantum Mechanical
  Observables}}.
\newblock {\slshape \bibinfo{journal}{Int.\ J. Theor.\ Phys.}}
  \bibinfo{volume}{{38}}, pp. \bibinfo{pages}{2441--2484},
  \doi{10.1023/A:1026628407645}.

\bibitemdeclare{incollection}{HC02}
\bibitem{HC02}
\bibinfo{author}{H.~\surnamestart Halvorson\surnameend} \&
  \bibinfo{author}{R.~\surnamestart Clifton\surnameend} (\bibinfo{year}{2002}):
  \emph{\bibinfo{title}{Reconsidering {Bohr's} reply to {EPR}}}.
\newblock In \bibinfo{editor}{T.~\surnamestart Placek\surnameend} \&
  \bibinfo{editor}{J.~\surnamestart Butterfield\surnameend}, editors: {\slshape
  \bibinfo{booktitle}{Non-locality and Modality}}, \bibinfo{publisher}{Kluwer},
  \bibinfo{address}{Dordrecht}, pp. \bibinfo{pages}{3--18},
  \doi{10.1007/978-94-010-0385-8_1}.

\bibitemdeclare{incollection}{How94}
\bibitem{How94}
\bibinfo{author}{D.~\surnamestart Howard\surnameend} (\bibinfo{year}{1994}):
  \emph{\bibinfo{title}{What makes a classical concept classical?}}
\newblock In \bibinfo{editor}{Jan \surnamestart Faye\surnameend} \&
  \bibinfo{editor}{Henry~J. \surnamestart Folse\surnameend}, editors: {\slshape
  \bibinfo{booktitle}{{Niels Bohr} and Contemporary Philosophy}},
  \bibinfo{publisher}{Kluwer}, \bibinfo{address}{Dordrecht}, pp.
  \bibinfo{pages}{201--229},
  \doi{10.1007/978-94-015-8106-6\_9}.

\bibitemdeclare{phdthesis}{How79}
\bibitem{How79}
\bibinfo{author}{D.~A. \surnamestart Howard\surnameend} (\bibinfo{year}{1979}):
  \emph{\bibinfo{title}{{Complementarity and Ontology: {Niels Bohr} and the
  Problem of Scientific Realism in Quantum Physics}}}.
\newblock Ph.D. thesis, \bibinfo{school}{Boston University}.

\bibitemdeclare{book}{Kal83}
\bibitem{Kal83}
\bibinfo{author}{G.~\surnamestart Kalmbach\surnameend} (\bibinfo{year}{1983}):
  \emph{\bibinfo{title}{Orthomodular Lattices}}.
\newblock \bibinfo{publisher}{Academic}, \bibinfo{address}{London}.

\bibitemdeclare{article}{KS67}
\bibitem{KS67}
\bibinfo{author}{S.~\surnamestart Kochen\surnameend} \& \bibinfo{author}{E.~P.
  \surnamestart Specker\surnameend} (\bibinfo{year}{1967}):
  \emph{\bibinfo{title}{The Problem of Hidden Variables in Quantum Mechanics}}.
\newblock {\slshape \bibinfo{journal}{J. Math.\ Mech.}} \bibinfo{volume}{17},
  pp. \bibinfo{pages}{59--87}, \doi{10.1512/iumj.1968.17.17004}.

\bibitemdeclare{article}{Mal77}
\bibitem{Mal77}
\bibinfo{author}{D.~\surnamestart Malament\surnameend} (\bibinfo{year}{1977}):
  \emph{\bibinfo{title}{Causal Theories of Time and the Conventionality of
  Simultaneity}}.
\newblock {\slshape \bibinfo{journal}{No\^{u}s}} \bibinfo{volume}{11}, pp.
  \bibinfo{pages}{293--300}, \doi{10.2307/2214766}.

\bibitemdeclare{article}{MS18}
\bibitem{MS18}
\bibinfo{author}{N.~D. \surnamestart Mermin\surnameend} \&
  \bibinfo{author}{R.~\surnamestart Schack\surnameend} (\bibinfo{year}{2018}):
  \emph{\bibinfo{title}{Homer Nodded: {Von Neumann's} Surprising Oversight}}.
\newblock {\slshape \bibinfo{journal}{Found.\ Phys.}} \bibinfo{volume}{48}, pp.
  \bibinfo{pages}{1007--1020}, \doi{10.1007/s10701-018-0197-5}.

\bibitemdeclare{book}{vN32E}
\bibitem{vN32E}
\bibinfo{author}{J.~\surnamestart von {Neumann}\surnameend}
  (\bibinfo{year}{1955}): \emph{\bibinfo{title}{{Mathematical Foundations of
  Quantum Mechanics}}}.
\newblock \bibinfo{publisher}{Princeton UP}, \bibinfo{address}{Princeton, NJ}.
\newblock \bibinfo{note}{[Originally published: {\it Mathematische Grundlagen
  der Quantenmechanik} (Springer, Berlin, 1932)]}.

\bibitemdeclare{article}{07TPQ}
\bibitem{07TPQ}
\bibinfo{author}{M.~\surnamestart Ozawa\surnameend} (\bibinfo{year}{2007}):
  \emph{\bibinfo{title}{Transfer Principle in Quantum Set Theory}}.
\newblock {\slshape \bibinfo{journal}{J. Symb.\ Log.}} \bibinfo{volume}{72},
  pp. \bibinfo{pages}{625--648}, \doi{10.2178/jsl/1185803627}.
\newblock \eprint{math/0604349}.

\bibitemdeclare{article}{14QST}
\bibitem{14QST}
\bibinfo{author}{M.~\surnamestart Ozawa\surnameend} (\bibinfo{year}{2014}):
  \emph{\bibinfo{title}{Quantum Set Theory Extending the Standard Probabilistic
  Interpretation of Quantum Theory (Extended Abstract)}}.
\newblock {\slshape \bibinfo{journal}{Electronic Proceedings in Theoretical
  Computer Science (EPTCS)}} \bibinfo{volume}{172}, pp.
  \bibinfo{pages}{15--26}, \doi{10.4204/EPTCS.172.2}.

\bibitemdeclare{article}{16A2}
\bibitem{16A2}
\bibinfo{author}{M.~\surnamestart Ozawa\surnameend} (\bibinfo{year}{2016}):
  \emph{\bibinfo{title}{Quantum Set Theory Extending the Standard Probabilistic
  Interpretation of Quantum Theory}}.
\newblock {\slshape \bibinfo{journal}{New Generat.\ Comput.}}
  \bibinfo{volume}{34}, pp. \bibinfo{pages}{125--152},
  \doi{10.1007/s00354-016-0205-2}.
\newblock \eprint{1504.06838}.

\bibitemdeclare{article}{17A1}
\bibitem{17A1}
\bibinfo{author}{M.~\surnamestart Ozawa\surnameend} (\bibinfo{year}{2017}):
  \emph{\bibinfo{title}{Operational Meanings of Orders of Observables Defined
  through Quantum Set Theories with Different Conditionals}}.
\newblock {\slshape \bibinfo{journal}{Electronic Proceedings in Theoretical
  Computer Science (EPTCS)}} \bibinfo{volume}{236}, pp.
  \bibinfo{pages}{127--144}, \doi{10.4204/EPTCS.236.9}.

\bibitemdeclare{article}{17A2}
\bibitem{17A2}
\bibinfo{author}{M.~\surnamestart Ozawa\surnameend} (\bibinfo{year}{2017}):
  \emph{\bibinfo{title}{Orthomodular-Valued Models for Quantum Set Theory}}.
\newblock {\slshape \bibinfo{journal}{Rev.\ Symb.\ Log.}} \bibinfo{volume}{10},
  pp. \bibinfo{pages}{782--807}, \doi{10.1017/S1755020317000120}.
\newblock \eprint{0908.0367}.

\bibitemdeclare{article}{21QTD}
\bibitem{21QTD}
\bibinfo{author}{M.~\surnamestart Ozawa\surnameend} (\bibinfo{year}{2021}):
  \emph{\bibinfo{title}{Quantum set theory: {Transfer Principle and De Morgan's
  Laws}}}.
\newblock {\slshape \bibinfo{journal}{Ann.\ Pure Appl.\ Log.}}
  \bibinfo{volume}{172}, pp. \bibinfo{pages}{102938/1--102938/42},
  \doi{10.1016/j.apal.2020.102938}.
\newblock \eprint{2002.06692}.

\bibitemdeclare{inproceedings}{21A6}
\bibitem{21A6}
\bibinfo{author}{M.~\surnamestart Ozawa\surnameend} (\bibinfo{year}{2021}):
  \emph{\bibinfo{title}{Reforming Takeuti's Quantum Set Theory to Satisfy de
  Morgan's Laws}}.
\newblock In \bibinfo{editor}{{Arai, T.} \surnamestart {\it et
  al.}\surnameend}, editor: {\slshape \bibinfo{booktitle}{Advances in
  Mathematical Logic}}, \bibinfo{publisher}{Springer Singapore},
  \bibinfo{address}{Singapore}, pp. \bibinfo{pages}{143--159},
  \doi{10.1007/978-981-16-4173-2_7}.
\newblock \eprint{2012.02928}.

\bibitemdeclare{article}{12RBR}
\bibitem{12RBR}
\bibinfo{author}{M.~\surnamestart Ozawa\surnameend} \&
  \bibinfo{author}{Y.~\surnamestart Kitajima\surnameend}
  (\bibinfo{year}{2012}): \emph{\bibinfo{title}{Reconstructing {Bohr's} Reply
  to {EPR} in Algebraic Quantum Theory}}.
\newblock {\slshape \bibinfo{journal}{Found. Phys.}} \bibinfo{volume}{42}, pp.
  \bibinfo{pages}{475--487}, \doi{10.1007/s10701-011-9615-7}.
\newblock \eprint{1107.0737}.

\bibitemdeclare{book}{Sak71}
\bibitem{Sak71}
\bibinfo{author}{S.~\surnamestart Sakai\surnameend} (\bibinfo{year}{1971}):
  \emph{\bibinfo{title}{{{C*}-Algebras} and {W*-Algebras}}}.
\newblock \bibinfo{publisher}{Springer}, \bibinfo{address}{Berlin}.

\bibitemdeclare{article}{Sas54}
\bibitem{Sas54}
\bibinfo{author}{U.~\surnamestart Sasaki\surnameend} (\bibinfo{year}{1954}):
  \emph{\bibinfo{title}{Orthocomplemented Lattices Satisfying the Exchange
  Axiom}}.
\newblock {\slshape \bibinfo{journal}{J. Sci. Hiroshima Univ. A}}
  \bibinfo{volume}{17}, pp. \bibinfo{pages}{293--302},
  \doi{10.32917/hmj/1557281141}.

\bibitemdeclare{article}{Sch35}
\bibitem{Sch35}
\bibinfo{author}{E.~\surnamestart {Schr\"{o}dinger}\surnameend}
  (\bibinfo{year}{1935}): \emph{\bibinfo{title}{{Die} {gegenw\"{a}rtige}
  {Situation} in der {Quantenmechanik}}}.
\newblock {\slshape \bibinfo{journal}{Naturwissenshaften}}
  \bibinfo{volume}{{23}}, pp. \bibinfo{pages}{807--812, 823--828, 844--849},
  \doi{10.1007/BF01491891, 10.1007/BF01491914, 10.1007/BF01491987}.
\newblock \bibinfo{note}{[English translation by J. D. Trimmer, Proc.\ Am.\
  Philos.\ Soc.\ {124}, 323--338 (1980)]}.

\bibitemdeclare{incollection}{Ta81}
\bibitem{Ta81}
\bibinfo{author}{G.~\surnamestart Takeuti\surnameend} (\bibinfo{year}{1981}):
  \emph{\bibinfo{title}{Quantum Set Theory}}.
\newblock In \bibinfo{editor}{E.~\surnamestart Beltrametti\surnameend} \&
  \bibinfo{editor}{B.~C. \surnamestart van Frassen\surnameend}, editors:
  {\slshape \bibinfo{booktitle}{Current Issues in Quantum Logic}},
  \bibinfo{publisher}{Plenum}, \bibinfo{address}{New York}, pp.
  \bibinfo{pages}{303--322}, \doi{10.1007/978-1-4613-3228-2_19}.

\end{thebibliography}
\end{document}